\documentclass[runningheads]{llncs}
\usepackage[T1]{fontenc}
\usepackage{graphicx}
\begin{document}
\title{Digital Persuasion: Understanding the Impact of Online Influencers on Public Opinion}

\author{Omran Berjawi\inst{1}\orcidID{0009-0008-6763-9630} \and
Rida Khatoun\inst{2}\orcidID{0009-0004-8282-136X} \and
Giuseppe Fenza\inst{3}\orcidID{0000-0002-4736-0113}}

\institute{IMT School for Advanced Studies Lucca, 55100 Lucca,
LU, Italy, 
\email{omran.berjawi@imtlucca.it}  
\and
Institut Polytechnique de Paris, 91120 Palaiseau, France 
\email{rida.khatoun@telecom-paris.fr} \and
University of Salerno, 84084 Fisciano, SA, Italy  
\email{gfenza@unisa.it}}

\maketitle       
\begin{abstract}
The studying of opinion dynamics and its propagation within social networks is crucial for addressing a wide range of challenges, including political polarization, public health, and marketing strategies. In this work, we study the problem of opinion dynamics by proposing a framework based on Friedkin-Johnsen (FJ) to identifies influential users and study their impact on dynamics opinions of community. The FJ model assume each individual have two opinions: initial and expressed. Through a series of initial opinion manipulation experiments, the proposed framework assesses the impact of influential versus random users on the overall community opinion. The proposed framework is validated using a tweet dataset representing the U.S. presidential election. The results shows that influencers with highest influencing score, significantly shift the overall community opinion. Moreover, the results shows that the impact of influencers not limited to direct neighbors , but beyond it, to their neighbors of neighbors . This study demonstrates how digital influencers on social media can shape public opinion regarding a subject or cause.

\end{abstract}

\section{Introduction}
\label{sec:introduction}
Social media platforms have grown tremendously and changed the way we communicate, exchange ideas, and form opinions. They have become a space through which any individual can express their opinions on any issue and anywhere in our world without any obstacles, which has led to their use as a sharpened tool in various aspects, including political discourse, consumer behaviors, and public health campaigns \cite{khanom2023using}. To understand the mechanisms of individuals’ opinion formation and how it is reshaped, it is significant to understand the relation between the intrinsic factors (e.g., pre-existing beliefs) and extrinsic influences (e.g., influencers, and media exposure). An important factor in this relationship is the role of influential users on reshaping opinions, and behaviors of their audience within their networks \cite{erikson2019american}. For instance, celebrities can play a significant role in public opinion during election campaigns. In the U.S. presidential election of 2020, figures like Dwayne "The Rock" Johnson and Taylor Swift allocated their platforms to promote voter participation and share their political beliefs. In 2024, Taylor Swift and Jennifer Lopez have publicly endorsed Kamala Harris, while Donald Trump has garnered support from prominent figures such as Elon Musk, Kid Rock, and Hulk Hogan.\

Various recent works have proposed various methods to identify influencers and study their relationship with opinion dynamics. For instance, the factors that influence the spread of opinion have been investigated in \cite{zhan2021bounded}. Other researchers investigated this relationship through optimizing opinion dynamic models \cite{sun2023opinion}. Some studies modify the user opinions in a graph \cite{zhou2023opinion}, while other approaches optimize the average opinion in the FJ \cite{gionis2013opinion}. More recent studies show the effectiveness of mathematical models in studying opinion dynamics within social networks. Among these models, the Friedkin-Johnsen (FJ) model \cite{friedkin1990social} has provided a foundational framework for studying opinion dynamics and has been confirmed by a sustained line of experiments \cite{zhou2024friedkin}. The FJ model simulates how opinions evolve through iterative interactions between individuals’ intrinsic beliefs and the expressed views of their neighbors. While the FJ model has been extensively applied in theoretical and empirical studies, its application to identifying influential users and simulating the impact of targeted opinion manipulation remains an area with significant potential for further exploration.\

In this context, we study the mechanisms of opinion dynamics within the social network graph by proposing an approach based on the FJ model. Specifically, this study focuses on the role of influencers and how manipulating their opinions can affect public opinion. By utilizing a publicly available dataset from Kaggle \cite{hui2020us}, which includes tweets collected during the U.S. presidential election of 2020, we demonstrate that interactions between certain influencers and various tweeters can change the opinions and convictions of the latter.
The primary contributions of this research are threefold:

\begin{itemize} \item Influencer identification: Identify and rank influential users within communities by interpreting the influence matrix from the FJ model at equilibrium.

\item Sentiment manipulation: Examine the impact of influencers on overall opinion by manipulating the initial opinions of influential users versus random users.

\item Propagation analysis: Analysis of the users most affected by the manipulation, who are directly related (neighbors) to the influencers or indirectly related (neighbors of neighbors).
\end{itemize}

This paper is organized as follows: In Section \ref{sec:related_work}, an overview of the existing literature on opinion dynamics models was discussed. Section \ref{sec:Methodology} details the approach used in this research. The experimental results are presented in Section \ref{sec:Experiments}. Finally, Section \ref{sec:Conclusion} summarizes the key contributions and concludes the paper.

\section{Related Works}
\label{sec:related_work}
The study on the opinion dynamic within online communities have become an hot topic, specifically, understand how opinion change and evolve between individuals. This section present the works the deal with these phenomena.\

Okawa et al. \cite{okawa2022predicting} proposed a hybrid method that integrates opinion dynamics models with deep learning techniques to enhance the predictive accuracy of opinion dynamics models. 
Zhou et al. \cite{zhu2020neural} used user past opinions, their context, neighbours opinions, as input to neural network to predict the user opinion. Other researches aimed in analyzing the role of influencers within social network. Arruda et al. \cite{de2024echo} study the how influencers behavior increase the intensity of echo chamber. In the same context, Galante et al. \cite{galante2023modeling} focused on how influencers' content spreading the emergence of echo chambers.\

In another line of research, other studies extend of the FJ model. For instance, Jia et al. \cite{tian2018opinion} examined the evolution of social influence within the FJ framework, highlighting how individuals' susceptibility to influence changes over time. \cite{bhalla2023local} proposed a extension of the FJ model, including a dynamic network structure in which connections are added or removed over time according to recommendations from the network. \cite{berjawi12024dynamic} utilized FJ model for analyzing how people’s opinions propagate through social networks and how influencers can affect these dynamics.

In same context, several studies employ the FJ model to analyze equilibrium opinions in social networks. Biondi et al. \cite{biondi2023dynamics} extract the conditions that FJ model induces opinion polarization in social networks by analyzing equilibrium opinions. In a similar vein, Zhou et al. \cite{zhou2024friedkin} interpreting equilibrium opinions of nodes through absorbing random walks to address challenges posed by signed relationships, such as the absence of doubly stochastic properties in the fundamental matrix.\

Zhang \cite{sun2023opinion} provides a valuable contribution by addressing the problem of opinion optimization in directed social networks using the FJ model. Their work focuses on minimizing and maximizing the average equilibrium opinion by modifying the internal opinions of selected K nodes to 0 and 1. Building on their approach, this work extends the use of the fundamental matrix $\Omega$. Specifically, the fundamental matrix was used to identify influential users whose opinions impact the equilibrium state of the network.
Our contribution complements their work by shifting the focus from optimization strategies to influencer identification and manipulation analysis. This approach provides a deeper understanding of the structural dynamics of opinion propagation and highlights the role of influencers in shifting sentiment changes by comparing the effects of targeted interventions at influential nodes to random interventions.

\section{Methodology}
\label{sec:Methodology}
This section describes in detail the employed methodology to study the opinion dynamics within social networks. The complete methodology is composed as follows: data preparation, graph construction \& community detection, and opinion dynamics modeling. The subsequent subsections outline each component in detail.

\subsection{Data Preparation}
\label{subsec:Data_Preparation}
First as preliminary data processing, the data used in this work consists of tweets shared by users within a social network. On Twitter, the interaction among users extend beyond static relationships (e.g., followers or friends), it consists of various content-based engagement such as mentions, replies, and retweets. The data preparation phase consists of the following steps:

\begin{itemize}

    \item Interaction Extraction: This step aims to extract the mention engagement within a tweet to represent the interaction between users, if one user mentions another user in their tweet, the frequency of mentions between users was calculated. These frequencies are associated as weights of interactions, reflecting the strength of influence between users.
    
    \item Sentiment Analysis: This steps focuses on measuring the user emotional stance on a specific topic through user tweets. For this purpose, the RoBERTa pre-trained model was employed to determine the sentiment polarity, ranging from negative to positive. The average sentiment score for each user was calculated and used as their initial opinion, providing a quantitative measure of the user’s overall stance.

\end{itemize}

\subsection{Graph Construction and Community Detection}  
\label{subsec:Graph_Construction}  

This component involves constructing a directed interaction graph, $G = (V, E)$, to represent user interactions. Here, $V = \{v_1, v_2, \cdots, v_n\}$ denotes the set of users, and $E$ represents the edges between users. The edges are created between two users if one mentioned the other in their tweets, with edge weights $a_{(v_i,v_j)}$ assigned as the frequency of mentions from $v_i$ to $v_j$. Once the network is constructed, the Louvain method \cite{blondel2008fast}, a modularity-based optimization algorithm, was employed to detect communities within the network.

\subsection{Opinion Dynamics Modeling}
\label{subsec:Opinion_Dynamics}

This phase represents the core of this research, which employs FJ model to simulate opinion dynamics. The FJ model assumes that each node in the network is characterized by two types of opinions: an initial opinion $s_i$ and an expressed opinion $x_i(t)$ at any given time $t$. The initial opinion $s_i$ represents the individual's inherent stance on a specific topic, expressed as a value in the range $[-1,1]$, where $s_i = -1$ indicates complete opposition to the topic, and $s_i = 1$ denotes full support. Importantly, $s_i$ remains constant throughout the process. In contrast, the expressed opinion $x_i(t)$ evolves over time, influenced by interactions with other users in the social network. The progression of these expressed opinions at $(t+1)$ is governed by the following equation:  

\begin{equation}  
x_i(t+1) = \frac{s_i + \sum_{j \in N(i)} a_{ij} x_j(t)}{1 + \sum_{j \in N(i)} a_{ij}} 
\end{equation}  

Here, $a_{ij}$ represents the weight of the interaction between $v_i$ and $v_j$, and $N(i)$ denotes the set of neighbors of $v_i$ .  

In this study, the FJ model was applied to the constructed graph. The average sentiment score (see Section \ref{subsec:Data_Preparation}) was used as $s_i$, while the weights of the connections between users were represented by $a_{ij}$.  The model iteratively updates each user’s $x_i(t)$ until the opinions converge, which means that each individual has stabilized into a fixed opinion. This process reaches a steady state represented by the equilibrium vector 
\begin{equation}
\mathbf{x} = (I + L)^{-1}.s 
\end{equation} 
where $I$ is the identity matrix, and $L$ is the Laplacian matrix.\

The matrix $(I + L)^{-1} $, also known as the fundamental matrix \cite{gionis2013opinion}, summarizes the structural relationships within the network, reflecting how the interactions and initial opinions of individuals  shape the final steady-state opinions. This formulation allows us to reveal how much initial opinion of each node influences other nodes' opinions and, conversely, how each node's opinion is influenced by the opinions of others in the network.

\subsubsection{Influence Identification}
\label{subsec:Influence_Identification}
We define influencers as those whose opinions significantly impact the opinions of others. This step is aimed at detecting and ranking the influential users by extracting their influence scores $InScore_i$ through the following equation:
\begin{equation}  
    InScore_i = \frac{1}{n} \sum_{j=1}^{n} \omega_{ij}  
\end{equation}  

where $\omega_{ij}$ are the fundamental matrix elements at the i-th row and the j-th column of the matrix.\

This score quantifies the influence of user $i$ on the equilibrium opinions of the network. A higher $InScore_i$ indicates that user $i$ has a high impact on the users final opinions. Using this score, users were ranked to identify key influencers within the community.

\subsubsection{Opinion Manipulation}
\label{subsec:FJ_optimization}
To study the impact of influencers on overall opinion,two sentiment manipulations was performed to assess how modifying the initial opinions of influential users, as compared to random users, affects the overall community sentiment. The two types of manipulation simulations are as follows:

\begin{itemize}

    \item Positive Manipulation: In this simulation, the initial opinions $s_i$ of the targeted user were set to $+1$, indicating a strongly positive sentiment.
    
    \item Negative Manipulation: In the second simulation, the initial opinions $s_i$ were set to $-1$, reflecting a strongly negative sentiment.

\end{itemize}

For each simulation, the FJ model was running to calculate a new set of equilibrium opinions, $x_i$. To study the changes in opinion of influencers versus random users, the average equilibrium opinion, denoted as \( Avg(z) \), was calculated, which capture the overall change in network sentiment.

\section{Experiments and Results}  
\label{sec:Experiments}  
This section detailed the experiments performed and present the result obtained from this work. First the influential users within community were identified and ranked. Following by studying their roles in shifting community opinion via manipulation their initial opinion. 

\subsection{Dataset}  
In this work, a publicly available dataset was employed from Kaggle \cite{hui2020us}, which contains more than 1.7 million tweets collected during the 2020 U.S. election and spans over one month. The dataset includes valuable metadata for each user, such as the number of followers, likes, retweets, and other engagement metrics. This dataset was selected due to its relevance in examining opinion dynamics within the context of a political event and allowing for a deeper analysis of user activity and interaction.

\subsection{Influencer Identification}
The influencer identification approach outlined in Subsection \ref{subsec:Influence_Identification} was applied in this experiment to detect the influential users. We ranked the top 15 influencers based on their $InScore$, and split them into three distinct batches:

\begin{itemize}  
    \item Batch 1: Top 5 influencers.  
    \item Batch 2: Influencers ranked 6–10.  
    \item Batch 3: Influencers ranked 11–15.  
\end{itemize}

\begin{table*}[ht]
\centering
\caption{Detected Influencers}
\label{tab:influencer_comparison}
\resizebox{\textwidth}{!}{%
\begin{tabular}{|c|l|c|c|c|c|c|}
\hline
\textbf{Rank} & \textbf{User} & \textbf{Batch} & \textbf{Influence Score} & \textbf{Retweets (Sum)} & \textbf{Number of Tweets} & \textbf{Followers Count} \\ \hline
1  & RealJamesWoods    & Batch 1 & 0.8908 & 109251 & 8  & 2,685,154 \\ \hline
2  & w\_terrence          & Batch 1 & 0.8805 & 42741  & 15 & 1,188,925 \\ \hline
3  & Varneyco        & Batch 1 & 0.8523 & 4800   & 55 & 663,854 \\ \hline
4  & Rasmussen\_Poll   & Batch 1 & 0.7605 & 2505   & 52 & 358,137 \\ \hline
5  & JudicialWatch     & Batch 1 & 0.7009 & 42773  & 87 & 1,843,739 \\ \hline
6  & EpochTimes    & Batch 2 & 0.5169 & 3232   & 36 & 319,635 \\ \hline
7  & trish\_regan   & Batch 2 & 0.4883 & 3152   & 6  & 737,555 \\ \hline
8  & WayneDupreeShow      & Batch 2 & 0.3430 & 1385   & 20 & 504,846 \\ \hline
9  & Styx666Official     & Batch 2 & 0.2417 & 417    & 6  & 90,474 \\ \hline
10 & realTrumpForce    & Batch 2 & 0.1662 & 12682  & 88 & 87,096 \\ \hline
11 & RealMattCouch        & Batch 3 & 0.1549 & 730    & 10 & 439,534 \\ \hline
12 & JoeTalkShow      & Batch 3 & 0.1159 & 12369  & 110 & 107,473 \\ \hline
13 & Wizard\_Predicts   & Batch 3 & 0.1155 & 1967   & 57 & 12,659 \\ \hline
14 & Out5p0ken    & Batch 3 & 0.1088 & 12     & 6  & 27,476 \\ \hline
15 & MarkSimoneNY    & Batch 3 & 0.1001 & 13579  & 97 & 193,717 \\ \hline
\end{tabular}%
}
\end{table*}

The purpose of distributing influencers users to several batches is to analyze the impact of $InScore$ on opinion dynamics. To validate the effectiveness of the proposed approach in identifying the influencers, we compared the results with the normalized centrality metric \cite{berjawi12024dynamic}, which is used to identify influencers.\ 

The results, as summarized in Table~\ref{tab:influencer_comparison}, detect the top 15 influencers within the network. It reveal that the FJ model successfully detected the same influential users as the normalized centrality metric method. Moreover, the FJ model shows its efficacy in ranking them by assigning $InScore$ to each user.\ 

\begin{figure*}
\centering
\includegraphics[width=\textwidth]{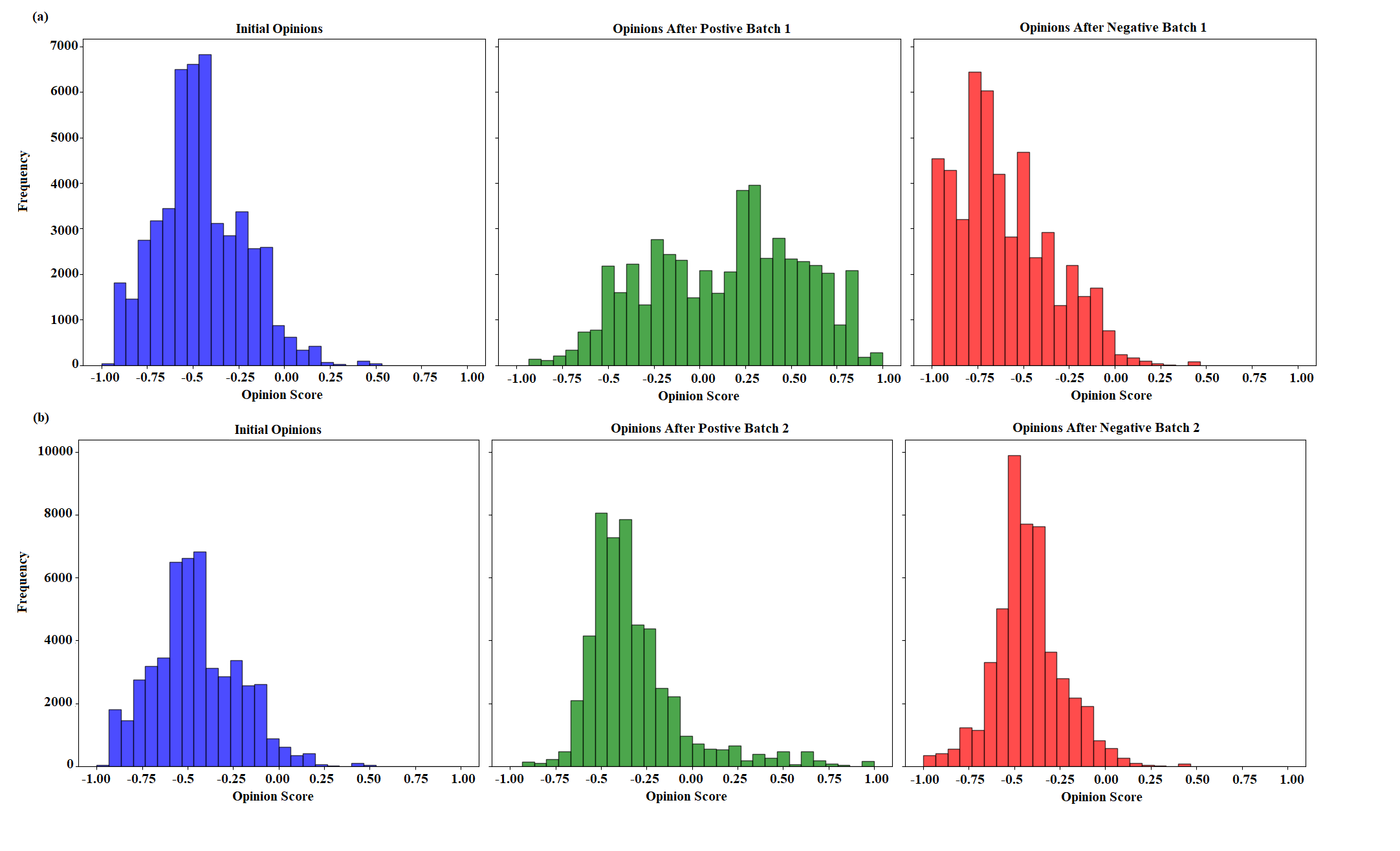}
\caption{A comparison of opinion distributions before and after positive and negative manipulation for Batch 1 and Batch 2.} \label{fig:batch1,2}
\end{figure*}

Additionally, the correlation between $InScore$ and the user engagements (total number of retweets, number of tweets, and follower count) of the identified influencers was investigated. As seen in Table~\ref{tab:influencer_comparison}, the correlation shows that there is a positive correlation between $InScore$ and Follower Count, as well as $InScore$ and Retweets (Sum). It highlights that influencers with higher follower counts and retweet engagements are associated with high $InScore$, while on the other hand, the number of tweets generated by influencers does not directly contribute to a user’s influence.\ 

These results indicate that some users engagement levels do not necessarily represent their influence, even if there is a large number of tweets. In contrast, some engagements, like the number of followers and retweets, have more meaning when studying their relation to the influence of the user.

\begin{figure*}
\centering
\includegraphics[width=\textwidth]{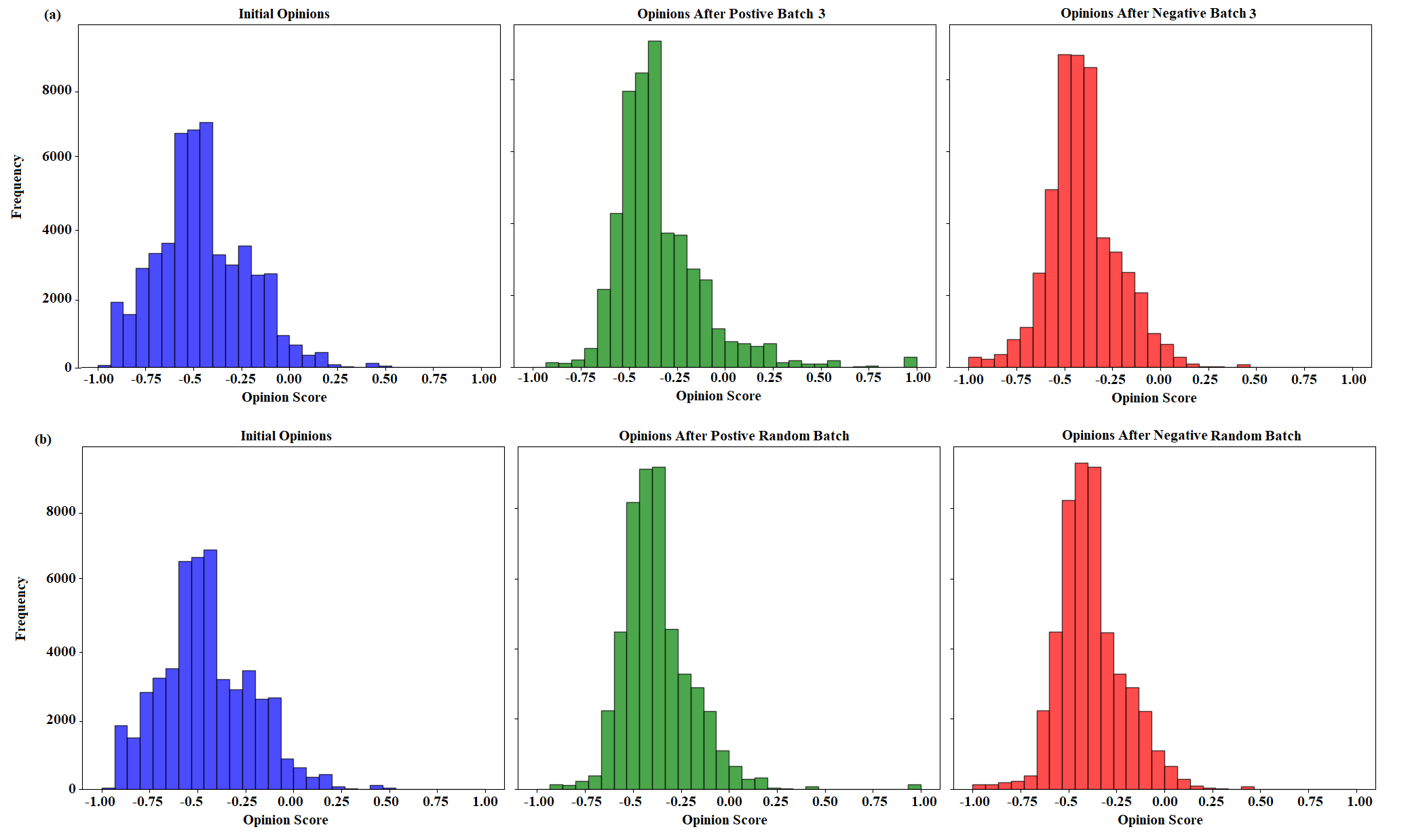}
\caption{A comparison of opinion distributions before and after positive and negative manipulation for Batch 3 and random Batch.} \label{fig:batch3,4}
\end{figure*}

\subsection{Sentiment Manipulation}
This experiment aims to investigate the impact of sentiment manipulation on the overall dynamic of community opinion. Specifically, we compare the opinion at equilibrium after simulating the positive and negative manipulation on the influential users versus random users.\ 

The results demonstrate that the top influencers (Batch 1) exerted a significantly stronger influence on the network's equilibrium opinion than lower-ranked influencers (Batch 2 and Batch 3) or random users. Specifically, as shown in Figure~\ref{fig:batch1,2}.a (top row) and Table~\ref{tab:opinion_changes}, positive sentiment manipulation on Batch 1 resulted in a significant rise in the average network opinion, from -0.3932 to 0.0866, whereas negative sentiment interventions amplified polarization by reducing the average opinion from -0.3932 to -0.5857.\ 

For Batch 2, the results in Figure~\ref{fig:batch1,2}.b show a rise in the average network opinion for positive manipulation, from -0.3932 to -0.3408, whereas negative sentiment manipulation reduces the average opinion from -0.3932 to -0.429. On the other hand, for Batch 3, as shown in Figure~\ref{fig:batch3,4}.a and Table~\ref{tab:opinion_changes}, there are minor changes in the average sentiment for both types of manipulation.\ 

Finally, the last batch, which represents the random users, shows a negligible impact on the overall sentiment. As depicted in Figures~\ref{fig:batch3,4}.b (bottom row), and Table~\ref{tab:opinion_changes}, for positive manipulation, the average opinion for both manipulation types provides negligible shifts in the network's average opinion, which suggests their limited roles. From -0.3932 to -0.3904 and from -0.3932 to -0.3940, respectively.\

Based on the above results, the findings support the hypothesis that influential users exhibit a clear ability to influence community opinion more effectively than other users, whether by amplifying positive emotions or exacerbating negative emotions. As observed, there is a strong relation between $InScore$ and the ability to shift the overall opinion of the community. When the $InScore$ increases, the opinion-shifting impact increases. Specifically, those with higher $InScores$ were able to induce significant changes in the community's sentiment, both in positive and negative directions. These results underline the importance of identifying and ranking influential users based on their influence score, as their influence score plays a crucial role in the impact on overall opinion of the community.

\begin{table} 
\centering
\caption{Equilibrium Opinion Changes by Batch}
\label{tab:opinion_changes}
\begin{tabular}{|c|c|c|c|}
\hline
\textbf{Batch} & \textbf{Original} & \textbf{Positive} & \textbf{Negative} \\ 
\textbf{} & \textbf{Opinion} & \textbf{Manipulation} & \textbf{Manipulation} \\ \hline
Batch 1 & -0.3932 & 0.0866 & -0.5857 \\ \hline
Batch 2 & -0.3932 & -0.3408 & -0.4291 \\ \hline
Batch 3 & -0.3932 & -0.3647 & -0.4120 \\ \hline
Random Batch & -0.3932 & -0.3904 & -0.3940 \\ \hline
\end{tabular}
\end{table}

\subsection{Propagation Analysis}

\begin{table}[]
\caption{Top 10 Affected Users for Positive/Negative Manipulation}
\label{tab:affected}
\begin{tabular}{|l|cc|cc|}
\hline
\textbf{}                      & \multicolumn{2}{c|}{\textbf{Direct Connection}}             & \multicolumn{2}{c|}{\textbf{Indirect Connection}}           \\ \hline
\textbf{}                      & \multicolumn{1}{l|}{Batch 1} & \multicolumn{1}{l|}{Batch 2} & \multicolumn{1}{l|}{Batch 1} & \multicolumn{1}{l|}{Batch 2} \\ \hline
\textbf{Positive Intervention} & \multicolumn{1}{c|}{7}       & 8                            & \multicolumn{1}{c|}{3}       & 2                            \\ \hline
\textbf{Negative Intervention} & \multicolumn{1}{c|}{6}       & 9                            & \multicolumn{1}{c|}{4}       & 1                            \\ \hline
\end{tabular}
\end{table}

After detecting the influencers and studying the relation between manipulating the initial opinion of influencers and its impact on the overall opinion of the community, we study how the influence propagate within the network. The propagation analysis investigates how opinion changes spread through the network through analyzing who the most affected users are by the opinion manipulation simulation.\ 
 
This experiment examined the top 10 most affected users for Batch 1 and Batch 2 influencers, who were identified as having the most substantial impact on the community’s collective sentiment, whether these users are directly connected to the influencers or influenced indirectly through neighbors of neighbors. As shown in Table~\ref{tab:affected}, we have seen a consistent pattern in the spread of sentiment changes for Batch 1 influencers. Both direct and indirect connections to the influencers are impacted by the two types of sentiment manipulation. The most impacted users are the influencers' neighbors; however, the impact also reached the indirect connections, demonstrating that the influencers can spread their influence through the network. On the other hand, for Batch 2, for both manipulations, most impacted users are the influencers' neighbors, in which 8 out of 10 for positive and 9 out of 10 for negative are the directed neighbors.\ 

In summary, regardless of the type of manipulation (positive or negative), the experiments provide valuable understanding into who is most affected by opinon manipulation. The findings validate that manipulating the initial opinion of the top influencer users (Batch 1) not limited to their neighbours, rather, it extends to neighbour of neighbours and, ultimately, to the broader community. In contrast, manipulating the initial opinion of Batch 2 influencers is more confined to their direct neighbors. These results suggest that as the influencer score increases, the influencer's impact expands beyond their immediate network to indirectly connected users and, eventually, to the overall network.

\section{Conclusion}
\label{sec:Conclusion}
This paper proposes a framework based on the FJ model to study the opinion dynamics within online communities. The proposed approach identifies the influential users and their impact on community opinion by analyzing the effects of manipulating their initial opinions, and examines how targeted manipulation of influential users can propagate through the network. The findings show that, in contrast to more traditional approaches, the suggested approach may identify influencers and rank them according to their influence score. The findings also highlight the fact that users with higher influence ratings have a significant impact on shifting opinions in the community and that their power extends beyond their immediate neighbors to include indirect relationships.\ 
In future research directions, the manipulation of opinion in real time may be investigated, in addition to validating the approach across different social network platforms.

\bibliographystyle{splncs04}
\bibliography{biblio}

\end{document}